\documentclass[11pt]{article}

\usepackage{amsthm,amssymb,verbatim,graphicx,subfigure,color,fullpage,times}

\newtheorem{lemma}{Lemma}
\newtheorem{theorem}{Theorem}

\newtheorem{myclaim}{Claim}

\newcommand{\csum}{{\Sigma}} % chromatic sum
\newcommand{\ecsum}{{\Sigma}'} % edge chromatic sum

\title{Minimum Sum Edge Colorings of Multicycles}

\author{Jean Cardinal\footnote{Universit\'e Libre de Bruxelles (ULB),
    CP 212. Supported by the Communaut\'e fran\c caise de Belgique -
    Actions de Recherche Concert\'ees (ARC). {\tt jcardin@ulb.ac.be}.}
  \and Vlady Ravelomanana\footnote{Laboratoire d'Informatique de
    l'Universit\'e de Paris-Nord (LIPN), 99 Av. J.-B. Cl\'ement, 93430
    Villetaneuse, France.} \thanks{\tt Vlady.Ravelomanana@lipn.univ-paris13.fr} \and 
    Mario Valencia-Pabon\footnotemark[2] \thanks{\tt valencia@lipn.univ-paris13.fr}}
\date{}

\begin{document}
\maketitle
\sloppy

\begin{abstract}
In the minimum sum edge coloring problem, we aim to assign natural 
numbers to edges of a graph, so that adjacent edges receive different numbers, and 
the sum of the numbers assigned to the edges is minimum. The {\em chromatic 
edge strength} of a graph is the minimum number of colors required in a minimum 
sum edge coloring of this graph. We study the case of multicycles, defined as cycles 
with parallel edges, and give a closed-form expression for the chromatic edge strength 
of a multicycle, thereby extending a theorem due to Berge. It is shown that the minimum sum can
be achieved with a number of colors equal to the chromatic index. We also propose simple algorithms
for finding a minimum sum edge coloring of a multicycle. Finally, these results are generalized
to a large family of minimum cost coloring problems.\\

\noindent {\bf Keywords}: graph coloring, minimum sum coloring, chromatic strength.
\end{abstract}

\section{Introduction}

{\em During a banquet, $n$ people are sitting around a circular table.
The table is large, and each participant can only talk to her/his
left and right neighbors. For each pair of neighbors around the table, 
there is a given number of available discussion
topics. If we suppose that each participant can only discuss one topic at
a time, and that each topic takes an unsplittable unit amount of 
time, then what is the minimum duration of the banquet, after which all
available topics have been discussed? What is the minimum average
elapsed time before a topic is discussed?}\\

In this paper, we show that there always exists a scheduling of the
conversations such that these two minima are reached simultaneously.
We also propose an algorithm for finding such a scheduling.
This amounts to {\em coloring} edges of a {\em multicycle} with $n$ vertices. \\

We first recall some standard definitions.
Let $G = (V,E)$ be a finite undirected (multi)graph without loops.  A
{\em vertex coloring} of $G$ is a mapping from $V$ to a finite
set of colors such that adjacent vertices are assigned different
colors. The {\em chromatic number} $\chi (G)$ of $G$ is the minimum
number of colors that can be used in a coloring of $G$.  An 
{\em edge coloring} of $G$ is a mapping from $E$ to a finite set of
colors such that adjacent edges are assigned different colors. The
minimum number of colors in an edge coloring of $G$ is called the 
{\em chromatic index} $\chi'(G)$.
From now on, we assume that colors are positive integers. The 
{\em vertex chromatic sum} of $G$ is defined as 
$\csum (G)= \min\left\{\sum_{v \in V}f(v) \right \}$, 
where the minimum is taken over all
colorings $f$ of $G$.  Similarly, the {\em edge chromatic sum} of $G$,
denoted by $\ecsum (G)$, is defined as 
$\ecsum (G) =\min\left\{\sum_{e \in E}f(e) \right \}$, 
where the minimum is taken over all
edge colorings. In both cases, a coloring yielding the chromatic sum is
called a {\em minimum sum coloring}.

We also define the minimum number of colors needed in a minimum sum
coloring of $G$.  This number is called the {\em strength} $s(G)$ of the graph $G$ 
in the case of vertex colorings, and the {\em edge strength} $s'(G)$ in
the case of edge colorings. Clearly, $s(G)\geq\chi (G)$ and
$s'(G)\geq\chi' (G)$.\\

The chromatic sum is a useful notion in the context of parallel job
scheduling. A {\em conflict graph} between jobs is a graph in which
two jobs are adjacent if they share a resource, and therefore cannot
be run in parallel. If each job takes a unit of
time, then a scheduling that minimizes the makespan is a coloring of
the conflict graph with a minimum number of colors. On the other hand,
a minimum sum coloring of the conflict graph corresponds to a
scheduling that minimizes the {\em average} time before a job is
completed. In our example above, jobs are conversations, resources are
the banqueters, and the conflict graph is the line graph of a
multicycle.

\paragraph{Previous results.} Chromatic sums have been introduced by
Kubicka in 1989~\cite{Kub89}.  The computational complexity of
determining the vertex chromatic sum of a simple graph has been
studied extensively since then. It is NP-hard even when restricted to some
classes of graphs for which finding the chromatic number is easy, such
as bipartite or interval graphs~\cite{BNK98,Sz99}. 
A number of approximability results for various classes of graphs were obtained in
the last ten years~\cite{barnoy98chromatic,GJKM02,HKS03,FLT04}.
Similarly, computing the edge chromatic sum is NP-hard for bipartite
graphs~\cite{GK00}, even if the graph is also planar and has maximum degree
3~\cite{marx-sum-edge-hard}. Hardness results were also given 
for the vertex and edge strength of a simple graph by 
Salavatipour~\cite{salavatipour03sum}, and Marx~\cite{marx-strength}.

Some results concern the relations between the chromatic 
number $\chi (G)$ and the strength $s(G)$ of a graph.
It has been known for long that the vertex strength can be arbitrarily
larger than the chromatic number~\cite{EKS90}. However, if $G$ is a proper interval
graph, then $s(G)=\chi (G)$ \cite{NSS99}, and $s(G)\leq\min \{ n, 2\chi (G) - 1\}$ if $G$
is an interval graph~\cite{Nicoloso04}. Hajiabolhassan, Mehrabadi, and Tusserkani~\cite{HMT00}  
proved an analog of Brooks' theorem for the vertex strength of simple graphs: 
$s(G)\leq \Delta (G)$ for every simple graph $G$ that is neither an odd cycle nor a
complete graph, where $\Delta (G)$ is the maximum degree in $G$.

Concerning the relation between the chromatic index and the edge
strength, Mitchem, Morriss, and Schmeichel~\cite{MMS97} proved an inequality 
similar to Vizing's theorem~: $s'(G)\leq\Delta
(G) + 1$ for every simple graph $G$. Harary and Plantholt~\cite{West} 
have conjectured that $s'(G)=\chi'(G)$ for every
simple graph $G$, but this was later disproved by Mitchem {\it et al.}~\cite{MMS97}, 
and Hajiabolhassan {\it et al.}~\cite{HMT00}.

\paragraph{Our results.} We consider {\em multigraphs}, in which
parallel edges are allowed. In Sections~\ref{sec:bip} and~\ref{sec:multicycles} 
we prove two main results:
\begin{enumerate}
\item if $G$ is a bipartite multigraph, then $s'(G)=\Delta (G)$,
\item if $G$ is an odd multicycle, that is, a cycle with parallel edges, of order $2k+1$ with $m$ edges, 
then $s'(G)=\max \{\Delta (G), \lceil m/k \rceil\}$.
\end{enumerate}
These statements extend two classical results from K\"onig and Berge, respectively.

In Section~\ref{sec:algorithms}, we give an algorithm of complexity $O(\Delta n)$ for finding a minimum sum 
coloring of a multicycle $G$ of order $n$ and maximum degree $\Delta$. This algorithm iteratively eliminates edges that will
form the color class corresponding to the last color $s'(G)$. For the special case where $n$ is even, we also give a more efficient 
$O(m)$-time algorithm based on the property of optimal colorings that the first color classes induce a uniform multicycle.

We conclude by generalizing our results to other objective functions.

\section{Bipartite Multigraphs}
\label{sec:bip}

The following well-known result has been proved by K\"onig in 1916.
\begin{theorem}[K\"{o}nig's theorem~\cite{K16}]
Let $G = (V,E)$ be a bipartite multigraph and let $\Delta$ denotes
its maximum degree. Then $\chi'(G) = \Delta$.
\end{theorem}
Hajiabolhassan{\it et al.}~\cite{HMT00} mention, without proof, that
$s'(G) = \chi'(G)$ for every bipartite graph $G$. We prove this result in 
the general case of bipartite multigraphs.\\

We first introduce some useful notations. 
Given $C$ the set of colors used in an edge coloring of a multigraph
$G$, we denote by $C_x$ the subset of colors of $C$ assigned to edges
incident to vertex $x$ of $G$. Given two colors $\alpha$ and $\beta$,
we call a path an $(\alpha,\beta)$-path if the colors of its edges 
alternate between $\alpha$ and $\beta$. We also denote by $d_G(x)$ the 
degree of vertex $x$ in $G$. 

\begin{theorem}
\label{thko}
Let $G = (V,E)$ be a bipartite multigraph and let $\Delta$ denote
its maximum degree. Then $s'(G) = \chi'(G) = \Delta$.
\end{theorem}

\begin{proof}
We proceed by contradiction. It is sufficient to assume that
there is a minimum sum edge coloring $f$ of $G$ using $\Delta +1$ colors. 
Let $C = \{1,\ldots,\Delta+1\}$ be the set of colors used by $f$. Choose
an edge $[a,b]_0$ of color $\Delta+1$. Clearly, $C_a \cup
C_b = \{1,\ldots,\Delta+1\}$, otherwise there exists a color $\alpha
\in \{1,\ldots,\Delta\}$ not used by any edge adjacent to both
vertices $a$ and $b$ that can be used to color edge $[a,b]_0$.
We would then obtain a new edge coloring $f'$ such that 
$\sum_{e \in E}f'(e) < \sum_{e \in E}f(e)$, a contradiction to the optimality of $f$.
Therefore, there exist colors $\alpha \in C_a \setminus C_b$ and
$\beta \in C_b \setminus C_a$ such that $\alpha, \beta \leq \Delta$.

Let $P_{\alpha\beta}$ denote a maximal $(\alpha,\beta)$-path
starting at vertex $a$.  Such a path cannot end at
vertex $b$, otherwise $G$ contains an odd cycle, contradicting the fact
that $G$ is bipartite. Thus we can recolor the edges of
$P_{\alpha\beta}$ by swapping colors $\alpha$ and $\beta$.  
After such a swap,  $\alpha \not \in C_a$ and $\alpha \not \in C_b$, 
hence we can assign color $\alpha \leq \Delta$ to $[a,b]_0$, obtaining a new
edge coloring $f'$. 

We prove that after such a recoloring,
\begin{equation}
\label{eq-decrease}
\sum_{e \in E}f'(e) < \sum_{e \in E}f(e).
\end{equation}
First, if $P_{\alpha\beta}$ has even length, the recoloring only changes
the color of $[a,b]_0$, hence~(\ref{eq-decrease}) holds. Otherwise, 
let $2s+1$ be the length of
$P_{\alpha\beta}$, with $s \geq 0$. Initially, the value of
the sub-sum corresponding to the
edge $[a,b]_0$ and to the $2s+1$ edges of $P_{\alpha\beta}$
in $f$ is $(\Delta + 1) + (s+1)\alpha + s\beta$ . After the
recoloring, the sub-sum in $f'$ has changed to 
$\alpha+ (s+1)\beta + s\alpha$.  The variation is $\beta - \Delta -1$. Since  
$\beta - \Delta -1 < 0$,  inequality~(\ref{eq-decrease}) holds, contradicting the 
optimality of $f$.  

Therefore, we have proved that if $f$ is an
edge coloring for $G$ such that $\sum_{e \in E}f(e) = \ecsum (G)$,
it uses at most $\Delta$ colors.
\end{proof}

\section{Multicycles}
\label{sec:multicycles}

Multicycles are cycles in which we can have parallel edges 
between two consecutive vertices. In this section we consider the 
chromatic edge strength of multicycles.\\

The chromatic edge strength $s'(G)$ of a graph $G$ is bounded from 
below by both $\Delta$ and $\lceil \frac{m}{\tau}\rceil$,
where $\Delta$ is the maximum degree in $G$ and $\tau$ is
the cardinality of a maximum matching in $G$. In this section, we show that 
the lower bound $\max\{\Delta, \lceil \frac{m}{\tau}\rceil\}$ is indeed 
tight for multicycles. We assume that the multiplicity of each edge in 
the multicycle is at least one, so that the size $\tau$ of a maximum 
matching is equal to $\lfloor n/2\rfloor$. In what follows, we let
$k=\lfloor n/2\rfloor$.

We first give a closed-form expression for the chromatic index of multicycles.
\begin{theorem}[\cite{B76}]
\label{thbeoriginal}
Let $G = (V,E)$ be a multicycle on $n$ vertices with $m$ edges and
degree maximum $\Delta$. Let $k=\lfloor n/2\rfloor$ denote the maximum cardinality of
a matching in $G$. Then
\begin{displaymath}
\chi'(G) =  \left\{\begin{array}{ll}
          \Delta, & \quad \textrm{if $n$ is even,}\\
          \max \left \{ \Delta, \lceil \frac{m}{k} \rceil\right \},
  & \quad \textrm{if $n$ is odd.} \\
          \end{array}\right.
\end{displaymath}
\end{theorem}

In order to determine the edge strength of a multicycle, we need the
following lemma proved by Berge in \cite{B76}.
\begin{lemma}[Uncolored edge Lemma \cite{B76}] 
\label{lebe}
Let $G$ be a multigraph without loops with $\chi'(G) = r+1$. If a coloring
of $G\setminus [a,b]_0$ using a set $C$ of $r$ colors cannot be extended to
color the edge $[a,b]_0$, then the following identities are verified : 
\begin{enumerate}
\item $|C_a \cup C_b| = r$,
\item $|C_a \cap C_b| = d_G(a) + d_G(b) - r - 2$,
\item $|C_a \setminus C_b| = r - d_G(b) + 1$,
\item $|C_b \setminus C_a| = r - d_G(a) + 1$.
\end{enumerate}
\end{lemma}

For proving Lemma~\ref{lebe}, we can solve the following linear system of $3$ equations on $3$ variables:
(i) $r = |C| = |C_a \cup C_b| = |C_a \cap C_b| + |C_a \setminus C_b| +
|C_b \setminus C_a|$; (ii) $|C_a \setminus C_b| = d_G(a) - 1 - |C_a
\cap C_b|$; and (iii) $|C_b \setminus C_a| = d_G(b) - 1 - |C_a \cap
C_b|$.\\

We now state our main result.
\begin{theorem}
\label{thbe}
Let $G = (V,E)$ be a multicycle on $n$ vertices with $m$ edges and maximum degree
$\Delta$, and let $k=\lfloor n/2\rfloor$ denote the maximum cardinality of a matching in $G$. Then
\begin{displaymath}
s'(G) = \chi'(G) =  \left\{\begin{array}{ll}
          \Delta, & \quad \textrm{if $n$ is even,}\\
          \max \left \{ \Delta, \lceil \frac{m}{k} \rceil\right \},
  & \quad \textrm{if $n$ is odd.} \\
          \end{array}\right.
\end{displaymath}
\end{theorem}

\begin{proof}
If $n$ is even, then the result follows from Theorem \ref{thko}. Thus, we
assume that $n = 2k+1$ for a positive integer $k$. We proceed by
induction on $m$, and let $r = \max \left \{ \Delta, \lceil\frac{m}{k}\rceil\right \}$.  

Assume that $m = 2k+1$. In this case, $G$ is a simple odd
cycle. Color the edges in $G$ in such a way that $k$ edges are
colored with color $1$, $k$ edges are colored with color $2$ and one
edge is colored with color $3$. Since $\Delta = 2$, $\lceil\frac{m}{k}\rceil=3$, 
and this coloring has minimum sum, the theorem holds for this case. 

Hence we assume that $m > 2k+1$ and, for the purpose of induction, 
that the result holds for all multicycles on $n$ vertices with fewer than $m$ edges. 
Let $[a,b]_0$ be an edge in $G$ and let $G' = G \setminus [a,b]_0$. By induction,
there exists a minimum sum edge coloring $f'$ of $G'$ using at most $r$ colors, 
with $r = \max \left \{ \Delta, \lceil \frac{m}{k} \rceil\right
\} \geq \max \left \{ \Delta', \lceil \frac{m-1}{k} \rceil\right
\} \geq \chi'(G')$ colors. Hence $G'$ is such that $s'(G') =
\chi'(G') \leq r$. 

Assume, by contradiction, that every minimum sum edge coloring $f$ of $G$ uses $r+1$
colors. The restriction of $f$ to edges in $G'$ 
must have minimum sum, otherwise contradicting the optimality of
$f$ in $G$. Hence the edge $[a,b]_0$ is the only edge in $G$ colored
by $f$ with color $r+1$. Let $C = \{1,\ldots,r\}$ be the set of
colors used by $f$ for the edges of $G'$ and, for each $i$ such that $1 \leq i \leq r$, 
let $E_i$ denotes the set of edges of $G'$ of color $i$. By induction, we have:
\begin{myclaim}
\label{cl2}
There exists a color $\sigma \in C$ such that $|E_\sigma| < k$.
\end{myclaim}
The claim holds, otherwise $m-1=\sum_{i=1}^r |E_i|=kr$, and
$r={m-1\over k}<\frac mk$, a contradiction.\\

\noindent By Lemma \ref{lebe}, $|C_a\cup C_b| = r$.
Hence it is sufficient to analyze the cases $\sigma \in
C_b \setminus C_a$ (or $\sigma \in C_a \setminus C_b$) and $\sigma \in
C_a \cap C_b$.\\

\begin{figure}[htp]
\begin{center}
\includegraphics[scale=.4]{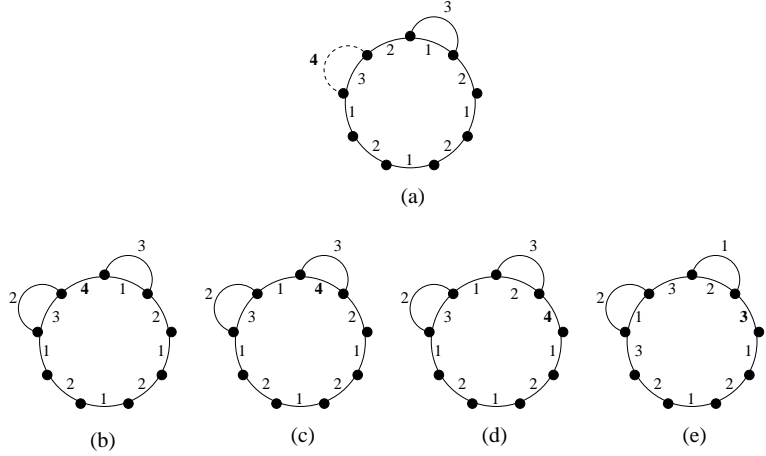}
\end{center}
\caption{(a) A multicycle $G$ where $\chi'(G) = 3$ and having an
edge colored with color $4$. In this example, $\sigma =3$. 
Figures~(a)-(c) illustrate the case $\sigma \in C_a \cap C_b$,
while Figures~(d)-(e) illustrate the case $\sigma \in C_b \setminus C_a$ in the proof
of Theorem \ref{thbe}.}
\label{mcyclefig}
\end{figure}

If $\sigma \in C_b \setminus C_a$, then by Lemma \ref{lebe},
there exists a color $\alpha \in Ca \setminus C_b$. Let
$G(\alpha,\sigma)$ denote the subgraph of $G'$ induced by the edges
of color $\alpha$ and $\sigma$. Let
$G_b(\alpha,\sigma)$ denote the connected component of
$G(\alpha,\sigma)$ containing the vertex $b$. Clearly,
$G_b(\alpha,\sigma)$ is a simple $(\sigma,\alpha)$-path having $b$ as
last vertex and not containing vertex $a$, otherwise we have a
contradiction to Claim~\ref{cl2}. Hence we can recolor the edges of
the path $G_b(\alpha,\sigma)$ by swapping colors $\alpha$ and
$\sigma$ in such a way that $\sigma \not \in C_b$. Since
$\sigma \not \in C_a$, we assign color $\sigma$ to $[a,b]_0$, and
obtain an edge coloring $f''$ of $G$ using $r$ colors. 
Figures~\ref{mcyclefig}~(d)-(e) provide an example of this case.

We now want to show that 
\begin{equation}
\label{eq-dec2}
\sum_{e \in E}f''(e) < \sum_{e \in E}f(e),
\end{equation}
contradicting $s'(G) > r$. If the length of the path
$G_b(\alpha,\sigma)$ is even, then $\sum_{e \in E}f''(e) - \sum_{e \in
  E}f(e) = \sigma - r - 1 \leq r - r - 1 < 0$.  If the length of the
path $G_b(\alpha,\sigma)$ is odd, say $2s+1$, with $s \geq 0$, then
the difference is $(\sigma + (s+1)\alpha + s\sigma) - (r+1
+ (s+1)\sigma + s\alpha) = \alpha - r -1\leq r - r - 1 < 0$.
Thus, inequality~(\ref{eq-dec2}) always holds.\\

The other case is when $\sigma \in C_a \cap C_b$. 
Then by Lemma \ref{lebe}, there exist colors $\alpha \in C_a \setminus C_b$ 
and $\beta \in C_b \setminus C_a$ with $\alpha \neq \beta \neq \sigma$. 
By induction, the result holds for $G' = G \setminus [a,b]_0$ and $G'$
has a minimum sum edge coloring using at most $r$ colors. 
Thus, the edge $[a,b]_0$ in $G$ is the only edge colored by $f$ with color 
$r+1$. 

Let us assume that vertices are ordered clockwise and let $b$ be the
right vertex of edge $[a,b]_0$.  Recolor edge $[a,b]_0$ with color
$\beta$ and the edge of color $\beta$ incident to $b$ with color $r+1$.
This recoloring does change neither the value of the sum nor the number of 
colors. Let $[x,y]_0$ be the edge that is recolored with color
$r+1$, with $x$ being its left vertex. 

By Lemma~\ref{lebe}, a color $\beta_y$ 
such that $\beta_y \in C_y \setminus C_x$ exists, otherwise there is 
a color $\theta \leq r$ such that $\theta \not \in
C_x$ and $\theta \not \in C_y$, and we can recolor $[x,y]_0$
with color $\theta$, contradicting the optimality of
$f$. We can therefore repeat the above procedure until the edge $[x,y]_0$ is such that 
$\sigma \in C_x \setminus C_y$
or $\sigma \in C_y \setminus C_x$. This is always possible,
because the cycle is odd, and $|E_\sigma| < k$. Assume,
without loss of generality, that $\sigma \in C_y \setminus C_x$. By
relabeling the vertices of $G$ in such a way that $x$ becomes $a$
and $y$ becomes $b$, we are back to the first case. 
Figures~\ref{mcyclefig}~(a)-(c) give an example of this case. 
\end{proof}

%%%%

\section{Algorithms}
\label{sec:algorithms}

We now present algorithms for minimum sum coloring of multicycles. Our algorithms assume that the encoding of the multicycle given as input has size $\Theta (n+m)$. This does not allow for implicit representations consisting of, for instance, the number of vertices and the number of parallel edges between each pair of consecutive vertices. This assumption is natural since we expect the resulting coloring to be represented by an encoding of size linear in the number of edges.

The line graph of a multicycle is a proper circular arc graph. Hence the problem of coloring edges of multicycles is a special case of proper circular arc graph coloring. It is easy to realize that not all proper circular arc graphs are line graphs of multicycles, though. Proper circular arc graphs were shown by Orlin, Bonuccelli, and Bovet~\cite{OBB81} to admit {\em equitable colorings}, that is, colorings in which the sizes of any two color classes differ by at most one, that only use $\chi$ colors. Therefore, a corollary of our results is that multicycles admit both equitable and minimum sum edge colorings with the same, minimum, number of colors, and that both types of colorings can be computed efficiently. 

We first present a general algorithm, then focus on the case where $n$ is even.

\subsection{The general case}

A natural idea for solving minimum cost coloring problems is to use a greedy algorithm that iteratively removes maximum independent sets (or maximum matchings in the case of edge coloring)~\cite{barnoy98chromatic,FLT04}. It can be shown that this approach fails here. Instead we use an algorithm in which the smallest color class, corresponding to color $s'$, is removed iteratively.\\

We first consider the case where $\lceil m / k\rceil \geq \Delta$ and $k$ divides $m$. Then the number of colors must be equal to $m/k$. But since each color class can contain at most $k$ edges, every color class in a minimum sum coloring must have size exactly $k$. Such a coloring can be easily found in linear time by a sweeping algorithm that assigns each color $i\bmod\chi'$ in turn. This is a special case of the algorithm of Orlin {\it et al.}~(Lemma~2,~\cite{OBB81}) for circular arc graph coloring. In the remainder of this section, we refer to this case as the "easy case".\\

\noindent {\bf Algorithm {\sc MulticycleColor}.}
\begin{enumerate}
\item $i\gets s' (G)$, $G_i\gets G$
\item \label{firsttest} {\bf if} $\lceil |E(G_i)| / k\rceil \geq \Delta (G_i)$ and $k$ divides $|E(G_i)|$ {\bf then}
apply the "easy case" algorithm and terminate
\item {\bf else}
\begin{enumerate}
\item \label{step1} Find a matching $M$ of minimum size such that $s' (G_i\setminus M)=s' (G_i)-1$
\item color the edges of $M$ with color $i$
\item $G_{i-1}\gets G_i\setminus M$, $i\gets i-1$
\item {\bf if} $G_i\not= \emptyset$ {\bf then} go to step~\ref{firsttest}
\end{enumerate}
\end{enumerate}

\noindent The correctness of the algorithm relies on the following lemma.
\begin{lemma}
\label{anymatching}
Given a matching $M$ in a multicycle $G$ such that 
\begin{enumerate}
\item\label{firstcondonM} $s' (G\setminus M)=s' (G)-1$, 
\item $M$ has minimum size among all matchings satisfying condition~\ref{firstcondonM},
\end{enumerate}
there exists a minimum sum edge coloring of $G$ such that $M$ is the set of edges colored with color $s' (G)$.
\end{lemma}
\begin{proof}
We distinguish three cases, a), b), and c), depending on the relative values of $\lceil m / k\rceil$ and $\Delta$.\\

{Case a)} We first assume that $\lceil m / k\rceil > \Delta$ and $k$ does not divide $m$, thus $m=\lfloor m/k\rfloor\cdot k+r$, with $r>0$. In that case, $M$ has size exactly $r$. To find a minimum sum coloring, we color the edges of $M$ with color $\lceil m / k\rceil$. The remaining edges are colored using the "easy case" algorithm, which applies since $\lfloor m/k\rfloor\geq\Delta$. This coloring must have minimum sum, because only one color class has not size $k$.\\

{Case b)} When $\Delta > \lceil m / k\rceil$, the matching $M$ is a minimum matching that hits all vertices of degree $\Delta$. We have to ensure that there exists a minimum sum coloring such that $M$ is the color class $s' (G)=\Delta$. 

We consider a minimum sum coloring and the color class $\Delta$ in this coloring. This class, say $M'$, must also be a matching hitting all vertices of degree $\Delta$. We now describe a recoloring algorithm that, starting with this coloring, produces a coloring whose sum is not greater and whose color class $\Delta$ is exactly $M$. We define a {\em block} as a maximal sequence of adjacent vertices of degree $\Delta$. The algorithm examines each block, and shifts the edges of $M'$ if they do not match with those of $M$. Two cases can occur, depending on the parity of the block length.

\begin{figure}
\begin{center}
\subfigure[\label{oddMprime}odd case: edges of $M'$]{\includegraphics[scale=.7]{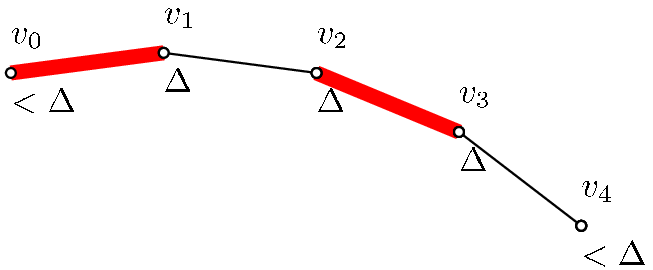}}\quad\quad 
\subfigure[\label{oddM}odd case: edges of $M$]{\includegraphics[scale=.7]{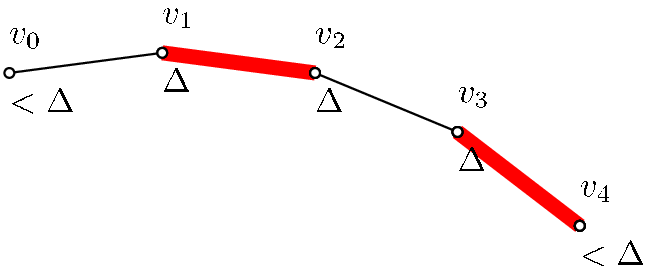}}\quad\quad\\

\subfigure[\label{evenMprime}even case: edges of $M'$]{\includegraphics[scale=.7]{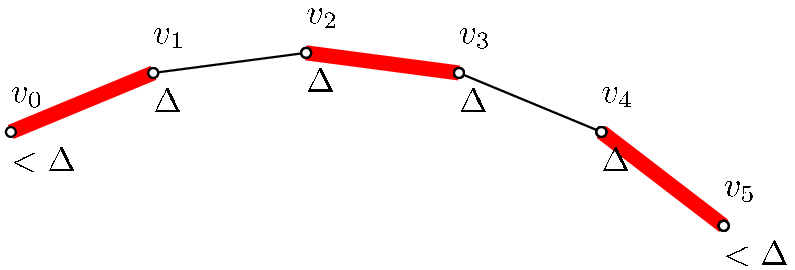}}\quad\quad
\subfigure[\label{evenM}even case: edges of $M$]{\includegraphics[scale=.7]{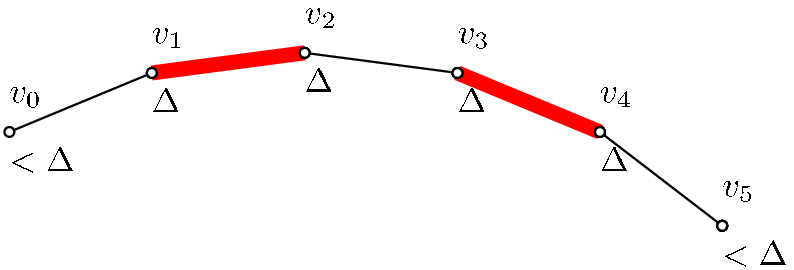}}
\end{center}
\caption{Illustration of the proof of Lemma~\ref{anymatching}.}
\end{figure}

The first case is when a block contains an odd number of vertices of degree $\Delta$, say $v_1, v_2, \ldots, v_{2t+1}$ for some integer $t$. In that case, the only way in which $M$ and $M'$ can disagree is, without loss of generality, when $M'$ contains edges of the form $v_0v_1, v_2v_3,\ldots ,v_{2t}v_{2t+1}$, while $M$ contains $v_1v_2, v_3v_4,\ldots ,v_{2t+1}v_{2t+2}$ (see figure~\ref{oddMprime}-\ref{oddM}), where $v_0$ and $v_{2t+2}$ are the predecessor of $v_1$ and the successor of $v_{2t+1}$, respectively. Since the degree of $v_0$ is, by definition of a block, strictly less than $\Delta$, there must exist a color $\alpha\in C_{v_1}\setminus C_{v_2}$. Furthermore, since all vertices within the block have degree $\Delta$, the color class for color $\alpha$ contains $t+1$ edges of the form $v_{2i+1}v_{2i+2}$ for $0\leq i\leq t$. Hence we can recolor the edges of $M'$ of the form $v_{2i}v_{2i+1}$ for $0\leq i\leq t$ with color $\alpha$, and the $t+1$ edges of color $\alpha$ within the block with color $\Delta$. Note that at this point, the coloring might be not proper anymore, as two edges colored $\Delta$ might be incident to $v_{2t+2}$. 

The other case is when a block contains an even number of vertices of degree $\Delta$, say $v_1, v_2, \ldots, v_{2t}$ for some integer $t$.
In that case, since $M$ is minimum, it contains edges of the form $v_1v_2, v_3v_4, \ldots, v_{2t-1}v_{2t}$. The only way in which $M'$ can disagree with $M$ is by containing edges $v_0v_1, v_2v_3,\ldots ,v_{2t}v_{2t+1}$ (see figure~\ref{evenMprime}-\ref{evenM}). Like in the previous case, there must be a color $\alpha\not\in C_{v_0}$, so we can recolor the edges of $M'$ of the form $v_{2i}v_{2i+1}$ for $0\leq i< t$ with color $\alpha$, and the edges of color $\alpha$ within the block with color $\Delta$. Note that at this point the coloring is not proper anymore, since the edges $v_{2t-1}v_{2t}$ and $v_{2t}v_{2t+1}$ both have color $\Delta$.

We proceed in this way for each block. Notice that the sum of the coloring is unaltered, and that the set of edges of color $\Delta$ is now a superset of $M$. Also, while $G$ is not necessarily properly colored anymore, the graph $G\setminus M$ is properly colored with at most $\Delta$ colors. But since removing $M$ decreases the strength, we know we can recolor $G\setminus M$ with $\Delta -1$ colors without increasing the sum. Doing that and glueing back the edges of $M$ colored with color $\Delta$, we obtain a minimum sum coloring where only the edges of $M$ have color $\Delta$, as claimed.\\

{Case c)} Finally, in the case where $\lceil m / k\rceil = \Delta$, with $m=\lfloor m/k\rfloor\cdot k+r$, the matching $M$ consists of at least $r$ edges that together hit all vertices of degree $\Delta$. If $M$ has exactly size $r$, then case a) above applies, since we know that by removing $M$, we also decrease the maximum degree. Otherwise case b) applies.
\end{proof}

We have to make sure that the main step of the algorithm can be implemented efficiently.

\begin{lemma}
Finding a matching $M$ in a multicycle $G$ such that $s' (G\setminus M)=s' (G)-1$ and $M$ has minimum size can be done in $O(n)$ time.
\end{lemma}
\begin{proof}
The three cases of the previous proof must be checked. In the case where $\lceil m / k\rceil > \Delta$ and $m=\lfloor m/k\rfloor\cdot k+r$, we can pick any matching of size $r$, which can clearly be done in linear time. In the second case, when $\lceil m / k\rceil < \Delta$, we need to find a minimum matching hitting all vertices of degree $\Delta$. This can be achieved in linear time as well by proceeding in a clockwise greedy fashion. 

Finally, in the last case, we need to find a minimum set of at least $r$ edges that together hit all vertices of degree $\Delta$. This can also be achieved in $O(n)$ time as follows. We first find the minimum matching hitting all maximum degree vertices. If the resulting matching has size at least $r$, then we are done and back to the previous case. Otherwise, we need to include additional edges. For that purpose, we can proceed in the clockwise direction and iteratively extend each block in order to include the exact number of additional edges. This can take linear time as well if we took care to count the size of each block and of the gaps between them in the previous pass.
\end{proof}

\begin{theorem}
Algorithm {\sc MulticycleColor} finds a minimum sum coloring of a multicycles with $n$ vertices and maximum degree $\Delta$ edges in time $O(\Delta n)$.
\end{theorem}
\begin{proof}
The number of iterations of the algorithm is at most $s'(G)=\max\{ \lceil m / k\rceil, \Delta\}$. Hence the running time is 
$O(\max\{ \lceil m / k\rceil n, \Delta n\})=O(\max\{m,\Delta n\})=O(\Delta n)$.
\end{proof}

We deliberately ignored the situation in which after some iterations,
the multicyle $G_i$ does not contain a full cycle anymore, that is, one of
the edge multiplicity $m_i$ drops to 0. We are then left with a
collection of disjoint {\em multipaths}, for which the minimum sum
coloring problem becomes easier. This special case is described in the
following section.

\subsection{A linear time algorithm for even length multicycles}

We turn to the special case $n=2k$, that is, the number of vertices is
even.  We show that in that case, minimum sum colorings have a
convenient property that can be exploited in a fast algorithm. This
algorithm first color a uniform multicycle contained in $G$ such that
the remaining edges of $G$ form a (possibly unconnected) multipath.
This multipath is then colored separately.\\

We begin this section by the following result on multipaths.
We consider multipaths with vertices labeled $\{1,2,\ldots ,n\}$, such that
edges are only between vertices of the form $i, i+1$.

\begin{lemma}
\label{lmpath}
There always exists a minimum sum edge coloring $f$ of a multipath $H$, 
such that its color classes $E_i$ are maximum matchings in the graphs 
$H'=H \setminus \cup_{j=1}^{i-1} E_j$; furthermore these matchings contain 
all the edges appearing in odd position from left to right in each connected 
component of $H'$.
\end{lemma}

\begin{proof}
Assume that $i>0$ is the minimum positive integer for which $E_i$ is
not a maximum matching of the graph $H' = H \setminus
\cup_{j=0}^{i-1} E_j$. Notice that the edges in $H'$ have a color at
least equal to $i$. Let $k$ be the size of a maximum matching in
$H'$.
  
We first suppose that $H'$ is connected. By hypothesis, we have that $|E_i|< k$.  
We consider a maximal sequence of consecutive vertices in $H'$, say 
$A = \{a_1,\ldots,a_{2t}\}$, such that one of the edges between vertices
$a_{2q-1}$ and $a_{2q}$ is colored by $f$ with color $i$, with $1 \leq q \leq t$.

If vertex $a_1$ is the second vertex in $H'$, then we
can recolor the edges of $H'$ as follows. 
Let $\alpha_0 \neq i$ be any color appearing on the edges 
between $a_0$ and $a_1$. We recolor
such an edge of color $\alpha_0$ with color $i$ and color the
edge between vertices $a_1$ and $a_2$ of color $i$ with color
$\alpha_0$. Now, for each $j$, with $1 \leq j < t$, we recolor the
edge $a_{2j+1}a_{2j+2}$ of color $i$ with a color $\alpha_j$
appearing on the edges $a_{2j}a_{2j+1}$, and color the edge
$a_{2j}a_{2j+1}$ of color $\alpha_j$ with color $i$.  The color
$\alpha_j$ is chosen such that $\alpha_j = \alpha_{j-1}$ if color
$\alpha_{j-1}$ appears on edges $a_{2j}a_{2j+1}$, and it is any color
appearing on edges $a_{2j}a_{2j+1}$ otherwise. 
At the end of this process, we have two cases.
(i) Color $\alpha_{t-1}$ appears on
the edges between vertices $a_{2t}$ and $a_{2t+1}$. In this case,
recoloring the edge $a_{2t}a_{2t+1}$ of color $\alpha_{t-1}$ with
color $i$, we obtain a new proper edge coloring of $H'$ whose sum
is less than the one induced by $f$, which is a contradiction. 
(ii) Color $\alpha_{t-1}$ does not appear on the edges
$a_{2t}a_{2t+1}$. In this case, recoloring any edge between vertices
$a_{2t}$ and $a_{2t+1}$ with color $i$, we obtain again a
contradiction to the optimality of $f$.  

If sequence $A$ begins at the $j$th vertex of $H'$, with $j>2$, we can assign color $i$ to an edge
between vertices $j-2, j-1$, and to an edge between vertices $j-4,j-3$, and so on, until the new 
sequence begins at the first or second vertex of $H'$. If it begins at the second vertex of 
$H'$, we apply the above recoloring procedure. Hence the sequence $A$ must begin at the first vertex of $H'$. 

If another, disjoint, such sequence follows sequence $A$ in $H'$, by using a similar recoloring 
argument, we can merge these two sequences, obtaining again a contradiction to the optimality 
of $f$. Therefore, $|E_i| = k$, and $E_i$ contains all the edges appearing at odd positions 
from left to right in $H'$.

Finally, if $H'$ is unconnected, the same reasoning can be applied to each connected 
component of $H'$.
\end{proof} 

From Lemma~\ref{lmpath}, we can deduce the following result, that settles the case of 
multipaths.
\begin{theorem}
\label{proppath}
The greedy algorithm that iteratively picks a maximum matching 
formed by all edges appearing in odd position in each connected component of
a multipath $H$, computes a minimum edge sum coloring of $H$ in time $O(m)$.
\end{theorem}

We now consider the case of even multicycles.
We assume that the vertices in the multicycle $G$ on $n = 2k$ vertices
are labelled clockwise with integers $0, 1, \ldots, n-1$, and
arithmetic operations are taken modulo $n$. For each $0 \leq i < n$,
we recall that $m_i$ denotes the number of parallel edges between two
consecutive vertices $i$ and $i+1$ in $G$.  Let $p$ be a positive
integer. A multicycle $G$ with $m = pn$ edges is called {\it $p$-uniform} 
if $m_i = p$ for every $i$ such that $0 \leq i < n$.

\begin{lemma}
\label{lmmv1}
Let $G$ be a multicycle of even length and let $p = \min_i{m_i}$.
Let $f$ be any minimum sum edge coloring of $G$. Then, $f$ can be
transformed into another minimum sum edge coloring $f'$ such that the
first $2p$ color classes $E_i$ induced by $f'$, with $1 \leq i \leq
2p$, are such that $|E_i| = k$ and their union induces a $p$-uniform multicycle.
\end{lemma}
\begin{proof}
Let $f$ be any minimum sum edge coloring of $G$, with $n=2k$ for some integer $k>1$. 
Clearly, as $f$ is minimum, we have that $|E_1| \geq |E_2| \geq \dots \geq |E_{\chi'}|$. 
Let us consider the following claim.
\begin{myclaim}
\label{clmv1}
The coloring $f$ can be transformed into a minimum sum edge coloring $f'$
having the property that the edges colored with colors $1$ and $2$
induce a subgraph of $G$ isomorphic to an cycle.
\end{myclaim}
\noindent Notice that, by using Claim \ref{clmv1}, the lemma follows directly by
induction on $p$. So, in order to prove Claim \ref{clmv1}, first
notice that, by using a similar recoloring argument as in the proof of
Lemma \ref{lmpath}, we can deduce that $|E_1| = k$. 

Now, without loss of generality, assume that $f$ is such that there is an 
edge colored with color $1$ between vertices $2j$ and $2j + 1$ for each 
$j$ with $0 \leq j < k$. Moreover, let $c \geq 2$ be the minimum color 
appearing on the edges between vertices $2j + 1$ and $2j + 2$, for all 
$0 \leq j < k$. 

Suppose that there exists a maximal sequence $i_1,\ldots,i_{2t}$ of consecutive
vertices in $G$, such that colors $1$ and $c$ belong to the set of
colors assigned by $f$ to the edges between vertices $i_{2q-1}$ and
$i_{2q}$, with $1 \leq q \leq t$. Then by using the same recoloring
argument as in the proof of Lemma~\ref{lmpath}, we can move color
$c$ in order to transform such a sequence into a $(c,1)$-path.
Moreover, again by using the same recoloring argument as in the proof
of Lemma~\ref{lmpath}, we can deduce that $|E_c| = k$. 

So, if $c = 2$ we are done, otherwise, we can swap the colors $2$ and $c$ 
so that $|E_2| = k$ and $E_1 \cup E_2$ induce a cycle.
\end{proof}

\begin{theorem}
There exists an $O(m)$-time algorithm for computing a minimum sum edge
coloring of $G$ of a multicycle of even length with $m$ edges. 
\end{theorem}
\begin{proof}
Let $n = 2k$ be the number of vertices in $G$ and let $p =\min_i\{m_i\}$, for 
$0 \leq i < n$. For each $0 \leq j < k$, assign to $p$ edges between vertices 
$2j$ and $2j+1$ the odd colors $1,3,\ldots,2p-1$ and assign to $p$ edges 
between vertices $2j+1$ and $2j+2$ the even colors $2,4,\ldots,2p$. 

The previous $pn$ colored edges induce a subgraph of $G$ isomorphic to a
$p$-uniform multicycle. When removing this $p$-uniform multicycle from $G$, we 
obtain a multipath or a set of disjoint multipaths, the edges of which can be colored
with colors in $\{2p+1,\ldots,s'(G)\}$, from Proposition~\ref{proppath}. 

Such a coloring can be computed in $O(m)$ time, and by 
Lemmas~\ref{lmmv1} and~\ref{lmpath}, it is a minimum sum edge coloring of $G$.
\end{proof}

%%%%

\section{Generalization}

In the generalized optimal cost chromatic partition
problem~\cite{jansen97}, each color has an integer cost,
but this cost is not necessarily equal to the
color itself. The cost of a vertex coloring is $\sum_{v\in V} c(f(v))$,
where $c(i)$ is the cost of color $i$. For any set of costs, our
proofs can be generalized to show that on one hand, the minimum 
number of colors needed in a minimum cost edge coloring of 
$G$ is equal to $\chi'(G)$ when $G$ is bipartite or a multicycle, 
and on the other hand that a minimum cost coloring can be 
computed in $O(\Delta n)$ time for multicycles.

In fact, our results can be generalized to a much larger class of
minimum cost edge coloring problems. 
Given an edge coloring $f:E\mapsto {\mathbb N}$, we define a cost 
$C(f)$ of the form:
$$
C(f) = \sum_i c(i, |f^{-1}(i)|),
$$
where $c:{\mathbb N}\times{\mathbb N}\mapsto {\mathbb R}$ is a 
real function of a color $i$ and an integer $k$, and $f^{-1}(i)$ is the set
of edges $e$ such that $f(e)=i$. Hence the cost to minimize is a sum of the cost
of each color class, itself defined as some function of the color and
the size of the color class. 

In the minimum sum coloring problem, the function $c$ is defined by
$$
c(i, k) = i\cdot k.
$$

We further suppose that the functions $c(i,k)$ satisfy the following property:

\noindent Given two nonincreasing integer sequences 
$a_1\geq a_2\ldots \geq a_n$ and $b_1\geq b_2\ldots \geq b_n$ such that
$$
\sum_{i=1}^j a_j \geq \sum_{i=1}^j b_j,
$$
we have 
\begin{equation}
\label{jungle}
\sum_{i=1}^n c(i, a_i) \leq \sum_{i=1}^n c(i, b_i).
\end{equation}

This property clearly holds in the minimum sum coloring problem.
It formalizes the fact that when minimizing the cost $C(f)$, we are looking for a distribution of the color class
sizes that is as nonuniform as possible. In particular, when an element (edge or vertex) in a color class $i$
is recolored with a color $j<i$, whose class is larger, then the objective function decreases. This is 
the argument that we implicitly used in our proof of Theorem~\ref{thbe}. It is also the argument that
ensures the correctness of the algorithms.

Property~(\ref{jungle}) can also be shown to hold (see~\cite{FHN08}) when the following two conditions
are satisfied:
\begin{enumerate}
\item $c(i,k)=c(j,k)\ \forall i,j$, that is, when the cost of a class only depends on its size, 
in which case we will say that the functions are {\em separable},
\item the functions $c(i,k)=c(k)$ are concave.
\end{enumerate}
This is the case for instance in the minimum entropy edge coloring problem~\cite{CFJ05}, 
for which $c(k)=-\frac km\log \frac km$. A number of other coloring problems falling in that class
were recently studied by Fukunaga, Halld\'orsson, and Nagamochi~\cite{FHN08}.\\

For all minimum cost edge coloring problems whose objective function satisfies~(\ref{jungle}), 
all our results apply. In fact, the colorings that we compute are {\em robust} colorings, in the sense that 
they minimize every objective function satisfying the above property.

\section*{Acknowledgments}

We thank Samuel Fiorini for insightful discussions on this topic.

\bibliographystyle{plain}

\begin{thebibliography}{10}
\bibitem{barnoy98chromatic}
A.~Bar-Noy, M.~Bellare, M.~M. Halld\'orsson, H.~Shachnai, and T.~Tamir.
\newblock On chromatic sums and distributed resource allocation.
\newblock {\em Information and Computation}, 140(2):183--202, 1998.
\bibitem{BNK98}
A.~Bar-Noy and G.~Kortsarz.
\newblock Minimum color sum of bipartite graphs.
\newblock {\em J. Algorithms}, 28(2):339--365, 1998.
\bibitem{B76}
C.~Berge.
\newblock {\em Graphs and Hypergraphs}.
\newblock North-Holland, 1976.
\bibitem{CFJ05}
J.~Cardinal, S.~Fiorini, and G.~Joret.
\newblock Minimum entropy coloring.
\newblock In {\em Proc. Int. Symp. on Algorithms and Computation ({ISAAC})}, 
volume 3827 of {\em Lecture Notes in Computer Science}, pages 819--828. Springer, 2005.
\bibitem{EKS90}
P.~Erd\"os, E.~Kubicka, and A.~Schwenk.
\newblock Graphs that require many colors to achieve their chromatic sum.
\newblock {\em Congressus Numerantium}, 71:17--28, 1990.
\bibitem{FLT04}
U.~Feige, L.~Lov{\'a}sz, and P.~Tetali.
\newblock Approximating min sum set cover.
\newblock {\em Algorithmica}, 40(4):219--234, 2004.
\bibitem{FHN08}
T.~Fukunaga, M.~Halldorsson, and H.~Nagamochi.
\newblock Robust cost colorings.
\newblock In {\em Proc. ACM-SIAM Symposium on Discrete Algorithms ({SODA})}, 2008 (to appear).
\bibitem{GJKM02}
K.~Giaro, R.~Janczewski, M.~Kubale, and M.~Malafiejski.
\newblock Approximation algorithm for the chromatic sum coloring of bipartite
  graphs.
\newblock In {\em Proc. Workshop on Approximation Algorithms for Combinatorial
  Optimization Problems ({APPROX})}, volume 2462 of {\em Lecture Notes in
  Computer Science}, pages 135--145. Springer, 2002.
\bibitem{GK00}
K.~Giaro and M.~Kubale.
\newblock Edge-chromatic sum of trees and bounded cyclicity graphs.
\newblock {\em Inform. Process. Lett.}, 75(1--2):65--69, 2000.
\bibitem{HMT00}
H.~Hajiabolhassan, M.~L. Mehrabadi, and R.~Tusserkani.
\newblock Minimal coloring and strength of graphs.
\newblock {\em Discrete Math.}, 215(1--3):265--270, 2000.
\bibitem{HKS03}
M.~M. Halld\'orsson, G.~Kortsarz, and H.~Shachnai.
\newblock Sum coloring interval and k-claw free graphs with application to
  scheduling dependent jobs.
\newblock {\em Algorithmica}, 37(3):187--209, 2003.
\bibitem{jansen97}
K.~Jansen.
\newblock Complexity results for the optimum cost chromatic partition problem.
\newblock In {\em Proc. Int. Conf. Automata, languages and programming
  ({ICALP})}, volume 1256 of {\em Lecture Notes in Computer Science}, pages
  727--737. Springer, 1997.
\bibitem{K16}
D.~K\"onig.
\newblock Gr\'afok \'es alkalmaz\'asuk a determin\'ansok \'es a halmazok
  elm\'elet\'ere.
\newblock {\em Matematikai \'es Term\'eszettudom\'anyi \'Ertes\'it\"o},
  34:104--119, 1916.
\bibitem{Kub89}
E.~Kubicka and A.~J. Schwenk.
\newblock An introduction to chromatic sums.
\newblock In {\em Proceedings of the {ACM} Computer Science Conf.}, pages
  15--21. Springer, 1989.
\bibitem{marx-sum-edge-hard}
D.~Marx.
\newblock Complexity results for minimum sum edge coloring, 2004.
\newblock Manuscript.
\bibitem{marx-strength}
D.~Marx.
\newblock The complexity of chromatic strength and chromatic edge strength.
\newblock {\em Comput. Complex.}, 14(4):308--340, 2006.
\bibitem{MMS97}
J.~Mitchem, P.~Morriss, and E.~Schmeichel.
\newblock On the cost chromatic number of outerplanar, planar, and line graphs.
\newblock {\em Discuss. Math. Graph Theory}, 17(2):229--241, 1997.
\bibitem{Nicoloso04}
S.~Nicoloso.
\newblock Sum coloring and interval graphs: a tight upper bound for the minimum
  number of colors.
\newblock {\em Discrete Mathematics}, 280(1-3):251--257, 2004.
\bibitem{NSS99}
S.~Nicoloso, M.~Sarrafzadeh, and X.~Song.
\newblock On the sum coloring problem on interval graphs.
\newblock {\em Algorithmica}, 23(2):109--126, 1999.
\bibitem{OBB81}
J.~B.~Orlin, M.~A.~Bonuccelli, and D.~P.~Bovet.
\newblock An $O(n^2)$ algorithm for coloring proper circular arc graphs.
\newblock {\em {SIAM}. J. Alg. Disc. Meth.}, 2(2):88--93, 1981.
\bibitem{salavatipour03sum}
M.~Salavatipour.
\newblock On sum coloring of graphs.
\newblock {\em Discrete Appl. Math.}, 127(3):477--488, 2003.
\bibitem{Sz99}
T.~Szkaliczki.
\newblock Routing with minimum wire length in the dogleg-free manhattan model
  is {NP}-complete.
\newblock {\em {SIAM} J. Comput.}, 29(1):274--287, 1999.
\bibitem{West}
D.~West.
\newblock Open problems section.
\newblock {\em The SIAM Activity Group on Discrete Mathematics Newsletter},
  5(2), Winter 1994--95.
\end{thebibliography}

\end{document}